\documentclass[aps,prb,twocolumn,floatfix,superscriptaddress,showpacs]{revtex4}

\usepackage{graphicx}
\usepackage{dcolumn}
\usepackage{bm}
\usepackage{textcomp}

\begin{document}

\title{Quantized coexisting electrons and holes in graphene measured using temperature dependent magneto-transport}

\author{E.~V.~Kurganova}
\email[]{E.Kurganova@science.ru.nl}
\affiliation{
High Field Magnet Laboratory, Institute for Molecules and Materials,
Radboud University Nijmegen, Toernooiveld 7, 6525 ED Nijmegen, The Netherlands
}

\author{S.~Wiedmann}
\affiliation{
High Field Magnet Laboratory, Institute for Molecules and Materials,
Radboud University Nijmegen, Toernooiveld 7, 6525 ED Nijmegen, The Netherlands
}
\author{A.~J.~M.~Giesbers}
\affiliation{
Department of Applied Physics, Eindhoven University of Technology, P.O. box 513, 5600 MB Eindhoven, The Netherlands}

\author{R.~V.~Gorbachev}
\affiliation{
School of Physics and Astronomy, University of Manchester, M13 9PL, Manchester, United Kingdom
}

\author{K.~S.~Novoselov}
\affiliation{
School of Physics and Astronomy, University of Manchester, M13 9PL, Manchester, United Kingdom
}
%

\author{M.~I.~Katsnelson}
\affiliation{Institute for Molecules
and Materials, Radboud University Nijmegen, Heyendaalseweg 135,
6525 AJ Nijmegen, The Netherlands }

\author{T.~Tudorovskiy}
\affiliation{Institute for Molecules
and Materials, Radboud University Nijmegen, Heyendaalseweg 135,
6525 AJ Nijmegen, The Netherlands }

\author{J.~C.~Maan}
\affiliation{
High Field Magnet Laboratory, Institute for Molecules and Materials,
Radboud University Nijmegen, Toernooiveld 7, 6525 ED Nijmegen, The Netherlands
}
\author{U.~Zeitler}
\email[]{U.Zeitler@science.ru.nl}
\affiliation{
High Field Magnet Laboratory, Institute for Molecules and Materials,
Radboud University Nijmegen, Toernooiveld 7, 6525 ED Nijmegen, The Netherlands
}

\begin{abstract}
We present temperature-dependent magneto-transport experiments around the charge
neutrality point in graphene and determine the amplitude of potential fluctuations $s$
responsible for the formation of electron-hole puddles. The experimental value $s \approx 20$~meV
is considerably larger than in conventional semiconductors which implies a strong
localization of charge carriers observable up to room temperature.
Surprisingly, in the quantized regime, the Hall resistivity overshoots the highest plateau values
at high temperatures. We demonstrate by model calculations that such a peculiar
behavior is expected in any system with coexisting electrons and holes when
the energy spectrum is quantized and the carriers are partially localized.
\end{abstract}


\maketitle

\section{Introduction}

Pristine graphene is a zero-gap semiconductor where the conduction
and valence band touch at zero energy. This point is generally referred
to as the charge neutrality point (CNP) with a zero charge carrier
density. However, it has been shown experimentally, that for
graphene on Si/SiO$_2$ substrates, electrons and holes coexist
around the CNP, which is attributed to the presence of
spatially inhomogeneous conducting electron-hole
puddles.\cite{TanPud,ZhangPud,puddlesMartin,Rutter}

In this paper we investigate the coexistence of electrons and holes by
means of temperature-dependent magneto-transport around the CNP.
We present results in the classical and quantum-Hall regimes.
In the classical regime, the individual charge carrier concentration and conductivity
at the CNP increases with increasing temperature due to thermal
activation but, as a consequence of potential fluctuations, do not vanish even at
the lowest temperature.  From this temperature dependence,
we determine the strength of the potential fluctuations responsible for
the formation of electron-hole puddles, $ s \approx 20$ meV for our samples.
In the quantum Hall regime, the Hall
resistance shows higher values than expected suggesting
a deactivation of charge carriers with
increasing temperature. Using model calculations, however,
we demonstrate that this counterintuitive temperature dependence
is straightforwardly explained considering the quantized density of states
of coexisting electrons and holes.

The paper is organized as follows: In Sec.~\ref{exper} we shortly describe
the samples an measurements. Sec.~\ref{class} is devoted to the experiential
magneto-transport in the classical regime along with an interpretation of
the results. In Sec.~\ref{quant} we  describe experimentally
(subsection ~\ref{quant}A) and discuss
theoretically (subsection ~\ref{quant}B) magneto-transport data in the quantum-Hall
regime. The paper ends with a conclusion in Sec.~\ref{concl}

\section{Experimental}
\label{exper}

We have measured two field effect transistors made from single
layer graphene (SLG) and bilayer graphene (BLG) deposited
on Si/SiO$_2$ wafers and shaped into a 1$~\mu$m wide Hall bar.
Both flakes originate from natural graphite and have mobilities
$\mu=1$~m$^{-2}$~s$^{-1}$ and $\mu=0.3$~m$^{-2}$~s$^{-1}$,
respectively. The total charge carrier concentration $n$
is controlled by applying a gate voltage $V_{g}$ on the conducting Si
substrate:\cite{Kostya2004}
\begin{equation}\label{nVg}
n=n_{e}-n_{h}=\alpha V_{g},
\end{equation}
with $\alpha= 7.2\times10^{14}$m$^{-2}$/V for a 300 nm thick SiO$_2$ gate insulator.
The subscripts $e$ and $h$ indicate electrons and holes, respectively.

Magneto-transport as a function of magnetic field, carrier concentration and temperature
were performed in a temperature range between 0.5 K and 250 K using
a top-loading He-3 system and a variable-temperature insert in a 33 T Bitter magnet.

\section{Classical Regime}
\label{class}

\begin{figure}[t!]
\includegraphics[width=\linewidth]{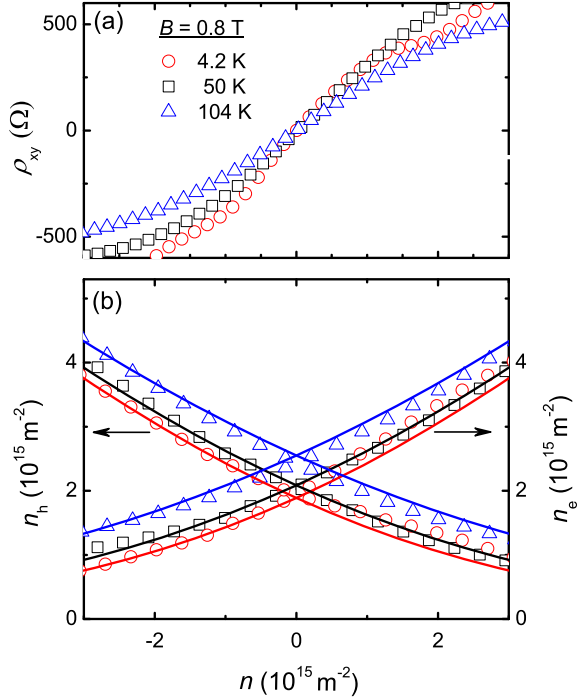}
\vspace*{-0.5cm}
\caption{(color online) (a) Hall resistivity of the BLG sample as a
function of the total charge carrier concentration at
$B=0.8$~T for various temperatures. (b) Extracted carrier
concentrations of electrons $n_{e}$ and holes $n_{h}$ for the same
field and temperatures (see text for details). The solid lines show
the calculated number of electrons and holes using the effective
density of states from Eq.~(\ref{DasDen}) and $s=19$~meV.}
\label{FigRxylowB}
\end{figure}

Fig.~\ref{FigRxylowB}(a) shows $\rho_{xy}(n)$ for the BLG sample at $B=0.8$~T for
different temperatures. Down to the lowest temperatures, the Hall resistivity
does not exhibit quantized plateaus at this magnetic field, indicating a
continuous energy spectrum. For all temperatures, the Hall
resistance smoothly crosses zero at the CNP, which indicates that electrons
and holes are present at all gate voltages.\cite{SteffenErik}

In this coexistence regime, the simultaneous contribution of electrons
and holes with equal mobility $\mu$ to the Hall effect is given by:\cite{Ashcroft, Das2cc}
\begin{equation}
\label{2carB}
\rho_{xy}=\frac{B}{e}\:\frac{(n_{e}-n_{h})(1+(\mu B)^{2})}{(n_{e}+n_{h})^{2}+((n_{e}-n_{h})\mu B)^{2}}.
\end{equation}
Note that this expression originally derived for massive charge carries with mobility $\mu=e \tau/m^*$ ($\tau$ is the elastic scattering time and $m^*$ is the effective mass) is also valid for massless particles when defining the mobility more generally as the ratio between the average drift velocity of the carriers and the applied electric field.

By solving the system of Eqs.~(\ref{nVg}) and (\ref{2carB}) with respect
to the variables $n_{e}$ and $n_{h}$ for each point of the experimental
$\rho_{xy}(V_{g})$ dependence, we can determine the number of electrons
$n_{e}$ and holes $n_{h}$ independently,\cite{SteffenErik} the results
are shown in Fig.~\ref{FigRxylowB}(b).

At the CNP ($n=0$) $n_e$ and  $ n_h $ are equal
to a residual concentration $n_0$, which is, according to Eq.~\ref{2carB},
related to the slope of $\rho_{xy}(n)$ at the CNP:
\begin{equation}\label{slope}
\left.\frac{d \rho_{xy}}{dn}\right|_{n=0}=\frac{B}{e}\:\frac{1+(\mu B)^{2}}{4n_{0}^{2}}.
\end{equation}

\begin{figure}[t!]
\includegraphics[width=\linewidth]{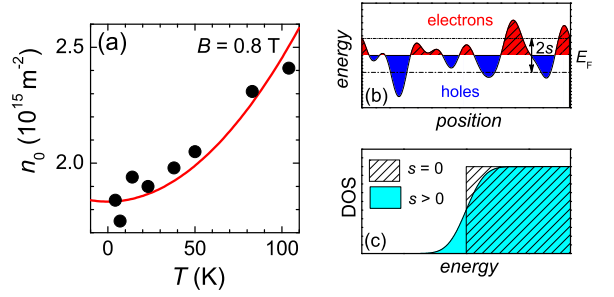}
\vspace*{-0.5cm}
\caption{(color online) (a) Electron and hole concentration $n_0$ of the BLG sample
at the CNP as a function of temperature.
The solid line represents a fit to Eq.~(\ref{Das}) with $s=19\pm2$~meV. \newline
(b) Sketch of the random potential fluctuations in the sample yielding a broadened
effective density of states shown in (c).\cite{DasS}}
\label{Fig-ne}
\end{figure}

The temperature dependence of $n_0$ is plotted in Fig.~\ref{Fig-ne}(a).
At zero temperature we find that a finite number of both electrons and holes
are present at the CNP. These electron-hole puddles originate from spatial
potential fluctuations within the sample represented schematically in
Fig.~\ref{Fig-ne}(b) and lead to
an effective density of states (DOS) around the CNP
plotted in Fig.~\ref{Fig-ne}(c):\cite{DasS}
\begin{equation}\label{DasDen}
D(E)=D_{0}\:{\rm erfc}(-E/\sqrt{2}s)/2,
\end{equation}
$s$ is the amplitude of potential fluctuations.

Though the question whether the puddles are caused by the impurity potential alone
or by intrinsic ripples in graphene is still under discussion,\cite{PoliniPud}
this phenomenological DOS is not sensitive to the deeper origin
of the puddles.
The calculated charge carrier concentrations using this effective DOS,
shown by the solid lines in Fig.~\ref{FigRxylowB}(b), indeed
reproduce the experimentally observed behavior around the CNP.

Directly at the CNP, the temperature dependence $n_0(T)$ in a low temperature ($k_{B}T/s<1$)
approximation is analytically found to be:\cite{DasS}
\begin{equation}
\label{Das}
n_{0}(T)=n_{0}(0)\left[1+\frac{\pi^{2}}{6}\left(\frac{k_{B}T}{s}\right)^{2}\right]
\end{equation}
and a fit of the data in  Fig.~\ref{Fig-ne}(a) to Eq.~\ref{Das}
yields $s=19\pm2$~meV.

\begin{figure}[t!]
\includegraphics[width=0.8\linewidth]{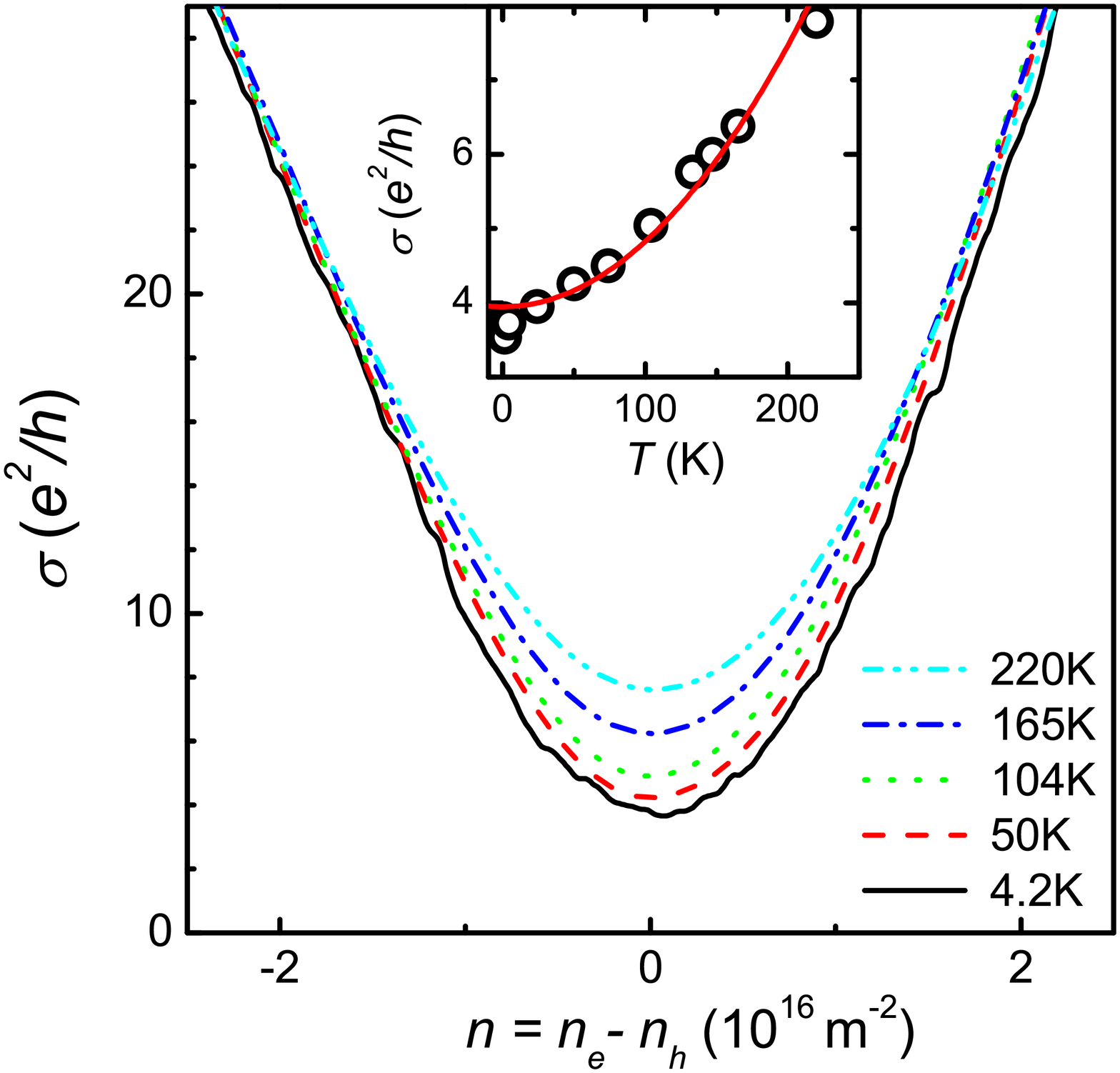}
\vspace*{-0.5cm}
\caption{(color online) Zero-field conductivity of the BLG sample around the CNP
for different temperature. The inset shows the temperature dependence of the minimum
conductivity at the CNP  with the solid line as a fit to Eq.~(\ref{DasSigma}) yielding $s=23\pm2$~meV.}
\label{Fig-sigma}
\end{figure}

Analogously, the temperature dependence of the conductivity
at the CNP can be described by:\cite{DasS}
\begin{equation}\label{DasSigma}
\sigma_{xx}(T)=\sigma_{xx}(0)\left[1+\sqrt{\frac{2}{\pi}}\frac{kT}{s}+\frac{\pi^{2}}{6}\left(\frac{k_{B}T}{s}\right)^{2}\right].
\end{equation}
The fit of the experimental temperature dependence in the classical regime
with Eq.~(\ref{DasSigma}) is shown in the inset Fig.~\ref{Fig-sigma} for the 0~T
trace. The fit yields $s=23\pm2$meV, which is close to the value $s=19\pm2$meV,
determined from the fit to the Hall data.

Remarkably, this strength of potential fluctuations is comparable to the thermal
energy at room temperature and considerably larger than in conventional semiconductor
heterostructures.\cite{Nixon} Therefore, whereas quantum Hall localization in GaAs-based systems
breaks down above 30 K,\cite{Rigal} the strong potential fluctuations can induce a robust
localization of charge carriers in graphene when the Fermi energy is situated between two Landau levels,
in particular around filling factor $\nu =2$.\cite{Giesbersgaps}
This allows the QHE to persist up to room-temperature.\cite{RTQH,Remark-InSb}

\section{Quantized Regime}
\label{quant}

\subsection{Experimental Results}

\begin{figure*}[t]
  \begin{center}
  \includegraphics[width=\linewidth]{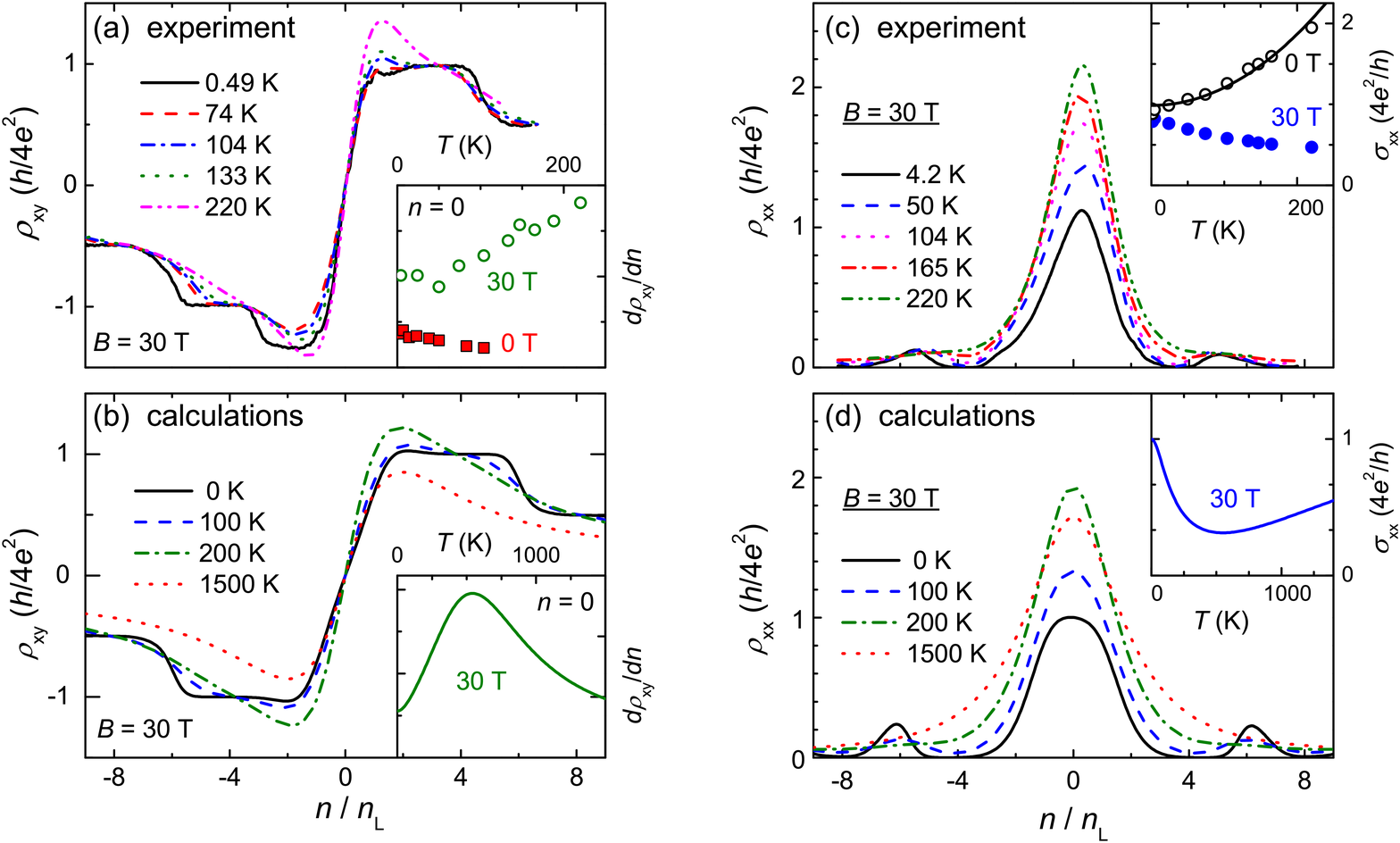}
  \end{center}
\vspace*{-1cm}
\caption{(color online) Transport coefficients $\rho_{xy}$ and $\rho_{xx}$ and of the BLG sample.
Panel (a) and (c) show the experimentally measured values which are compared to
the calculated quantities in (b) and (d). The insets in (a) and (c)
show the corresponding slopes of the Hall resistivity at the CNP as a function of temperature.
The insets in (b) and (d) show the  temperature dependance of the conductivity at the CNP.}
\label{FigAllR}
\end{figure*}

Fig.~\ref{FigAllR}(a)
shows the measured Hall effect for BLG as a function of the total
charge carrier concentration $n$ at $B=30$~T for various temperatures.
For $B\geq15$~T the system is in the quantized regime and we observe plateaus
at $\nu=\pm4,\pm8\dots$ as expected for BLG. At the
lowest temperatures, also a plateau develops at $\nu=-3$ which makes
the $\rho_{xy}$ curve slightly asymmetric.
However, since the plateau is smoothed out for $T>74$~K, we neglect this asymmetry
in the following.

At $T=0.5$~K, the Hall resistance on the
electron (hole) side reaches its maximum (minimum) value at the plateau
and goes smoothly to the hole (electron) side as already observed in the non-quantized
regime and this behavior can again be qualitatively explained by the two-carrier model.
At high temperatures, the Hall resistivity overshoots the plateau $\nu=\pm4$ before the beginning
of the plateau and reaches values higher than $h / 4e^{2}$ and $\rho_{xy} (n)$ becomes steeper at
the CNP, see Fig.~\ref{FigAllR}(a).
We observe a similar behavior in the SLG sample, therefore,
this effect has a generic nature and was also observed by others.\cite{Kimoblom}
Note that these overshoots in the Hall resistance only appears at high temperatures and is
therefore not due to the (partial) splitting of the lowest Landau level as observed in e.g.~
Ref.~\onlinecite{Kimnu0}.

The temperature dependence of the slope of $\rho_{xy} (n)$ at the CNP
is explicitly shown in the inset to Fig.~\ref{FigAllR}(a).
It is opposite to the classical regime, which can
be seen by comparison of the traces at $B=30$~T (quantized regime)
and $B=0.8$~T (classical regime), plotted together in the inset.

According to Eq.~(\ref{slope}), an increase in the slope with increasing temperature
corresponds to a decrease of the charge carrier concentration, which
appears to be counter-intuitive.
However, the plateaus remain visible up to the highest temperature measured,
i.e., the system remains quantized and the number of electrons and holes
within one Landau level is fixed. In the quantized regime, the slope of the
Hall resistivity depends not only on the number of charge carriers at the CNP,
but also the strength of localization, i.e.~the width of the plateaus closest
to the CNP.
Therefore, Eq.~(\ref{slope}) cannot be used to relate the slope to
the number of charge carriers.
In fact, Eq.~(\ref{2carB}), which predicts a $1/n$-behavior of the slope around
the CNP when $\mu B\rightarrow\infty$, anticipates such an overshoot.

A similarly unexpected behavior for high temperature is observed in the
resistance at high magnetic fields. Fig.~\ref{FigAllR}(c) shows the
resistivity of BLG in the quantized regime as a function of the total charge
carrier concentration for various temperatures. In contrast to the non-quantized
regime, the resistivity exhibits a metal-like increase with temperature.

As we will show in the following subsection, this peculiar
transport properties in the quantized regime can be described in a
simple DOS model with localized and extended states where
the temperature dependence of the resistivity is governed
by a redistribution of charge carriers due to the thermal activation.

\subsection{Model Calculations and Discussion}

\begin{figure}[t!]
  \includegraphics[width=0.7\linewidth]{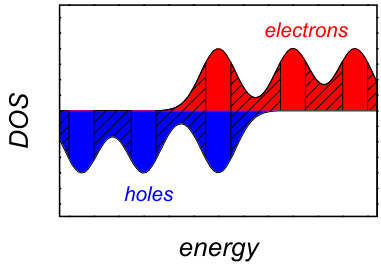}
\vspace*{-0.5cm}
\caption{(color online) Schematic density of states (DOS) of broadened Landau levels
in BLG. The dashed regions correspond to localized states between two levels.}
\label{FigDOS}
\end{figure}

The model used for our calculation starts with the quantized DOS for graphene in magnetic field consists of separate
Gaussian-shaped Landau levels (LLs) as depicted schematically in Fig.~\ref{FigDOS}.
States in the middle of each LL are
extended, and states are localized in the tails.
For our simulations of the DOS in BLG shown below, we have
used a constant width $\Delta = 100$~K for the higher Landau
levels and $\Delta = 200$~K for the zero-energy LL.\cite{Bilayergaps}
Landau levels above (below)
the CNP, populated by electrons (holes), have a degeneracy of $4eB/h$. The
zeroth LL with a degeneracy of $4eB/h$ for SLG and $8eB/h$ for BLG
is populated by electrons and holes simultaneously,
such that electron and hole conduction can be found both above and
below the CNP. We can then calculate the conductivity, using the
Kubo-Greenwood formalism,\cite{Kubo,Greenwood}:
\begin{eqnarray}\label{sigmaxx}
\sigma_{xx} & = & \sigma_{xx}^{e}+\sigma_{xx}^{h}\\
         & = &\int_{-\infty}^{\infty}\left[ K^{e} (E) \frac{\partial f(E)}{\partial E} +
         K^{h} (E) \frac{\partial(1-f(E))}{\partial E}\right]\,\mathrm{d}E\nonumber.
\end{eqnarray}
Here $f(E)$ is the Fermi distribution and $K^{e,h} (E)$ is an energy dependent function
containing the density of the extended states, the particle velocity and the scattering time.
the upper indices $e,h$ correspond to electrons and holes, respectively.
Modeling $K(E)$ by a superposition of Gaussians with the above mentioned widths,
the Kubo-Greenwood formula reproduces qualitatively the behavior
of the experimental conductivity in graphene, but does not give a universal
value of minimal conductivity.\cite{Kubononun} Therefore, we normalized the
integral in Eq.~(\ref{sigmaxx}) such that, at zero temperature,
$\sigma_{xx}$ at the CNP has the universal value of $4e^{2}/h$ in accordance
with the experimentally observed value.

Similarly, we can determine the Hall conductivity by summing up all extended
states below the  Fermi level, smeared out by temperature:
\begin{equation}\label{sigmaxy}
\sigma_{xy}\propto \int_{-\infty}^{\infty}\left[D^{h}_{ext}(1-f(E))-D^{e}_{ext} f(E)\right]\,\mathrm{d}E
\end{equation}

When introducing a localized DOS,
the calculated conductivity  $\sigma_{xy}$ develops plateaus which are not
at multiples of $4e^2/h$,\cite{vonK1985} because the
number of extended states within a LL is less than the total
level degeneracy.  In order to reproduce the correct number of occupied
LLs in Eq.~(\ref{sigmaxy}), we have normalized the extended DOS for each LL
to the total number of extended states within it.

After calculating the conductivities we convert them into
the resistivity $\rho_{xx}$ and the Hall resistivity
$\rho_{xy}$  using the standard matrix relations, the results
are shown in Fig.~\ref{FigAllR}(b) and (d).
They can be compared to the experimental
curves, plotted as panels (a) and (c) of the figures.
The calculated curves reproduce qualitatively the experimental
results, and display a similar temperature dependence.
In particular, at the CNP,
the resistivity (insets in Fig.~\ref{FigAllR}(c) and (d))
and the slope of the Hall resistivity
(insets in Fig.~\ref{FigAllR} (a) and (b))
increase with increasing temperature up to 200~K.
Moreover, as is also seen in the experiment, the overshoots
develop at the beginning of the plateaus at high temperatures
(compare panels (a) and (b) in Fig.~\ref{FigAllR}).

Starting from about 500~K (as calculated for 30~T) the system smoothly enters the
classical regime. The conductivity (inset to Fig.~\ref{FigAllR}(d))
starts to increase with increasing temperature and the slope of the Hall resistivity
(inset to Fig.~\ref{FigAllR}(b)) decreases.
Experimentally, this trend in the non-quantized regime is verified
by the low field data presented earlier in Figs.~\ref{Fig-ne} and \ref{Fig-sigma}.

\section{Conclusions}
\label{concl}

To conclude, we have investigated the coexistence of electron and hole
magneto-transport in graphene for a wide range of temperatures and
magnetic fields. For all temperatures and fields, the Hall resistivity
smoothly crosses zero at the CNP indicating the simultaneous presence of
both electrons and holes at the CNP. In the non-quantized regime, the
slope of the Hall resistivity decreases with increasing temperature,
which is accompanied by an increasing conductivity at the CNP.
This  behavior can be quantitatively modeled by a thermal
activation of charge carriers experiencing relatively large
potential fluctuation of the order of 20 meV.

In the quantum Hall regime, increasing the temperature has the opposite effect on
the slope of the Hall effect and produces counter-intuitive overshoots
when  approaching the $\nu = \pm 2$ plateaus.
Using an appropriate density of states, we
have shown in model calculations that such a behavior is expected
for any two-carrier system with a quantized energy spectrum.

\section{Acknowledgments}

Part of this work has been supported by EuroMagNET under EU contract
RII3-CT-2004-506239 and by the Stichting Fundamenteel Onderzoek
der Materie (FOM) with financial support from the Nederlandse
Organisatie voor Wetenschappelijk Onderzoek (NWO).



\begin{thebibliography}{23}
\bibitem [{Rem()}]{Remark-InSb}%
  \BibitemOpen
  \href@noop {} {}\bibinfo {note} {The recent observation of a robust quantum
  Hall effect in InSb systems (Ref.~\onlinecite{Alexander-Webber}) may indicate
  that such a robust localization can also be achieved in conventional
  semiconductor heterojunctions.}\BibitemShut {Stop}%
\end{thebibliography}

%

\end{document}